\documentstyle[aps,preprint,prb,psfig]{revtex}
\begin{document}
\draft
\title{Mobility and diffusivity in a generalized
Frenkel-Kontorova model.}
\author{Oleg~M.~Braun~$^{\dag,\S}$, Thierry~Dauxois~$^{\S}$ ,
Maxim~V.~Paliy~$^{\dag}$, and Michel~Peyrard~$^{\S}$}

\address{$^{\dag}$ Institute of Physics, the Ukrainian National Academy
of Sciences, 46 Science Avenue, UA-252022 Kiev, Ukraine}

\address{$^{\S}$ Laboratoire de Physique, Ecole Normale
Sup\'{e}rieure de Lyon, 46 All\'{e}e d'Italie, 69364 Lyon
C\'{e}dex 07, France}
\date{\today}

\maketitle

\begin{abstract}

Molecular dynamics simulations are used to investigate
the atomic mobility and diffusivity of a generalized
Frenkel-Kontorova model which takes into
account anharmonic (exponential) interaction of atoms
subjected to a three-dimensional substrate potential
periodic in two dimensions and nonconvex (Morse) in the
third dimension. The numerical results are explained by a
phenomenological theory which treats a system of strongly
interacting atoms as a system of weakly interacting
quasiparticles (kinks). Model parameters are chosen
close to those for {\bf K}-{\bf W}(112) adsorption system.

\end{abstract}
\bigskip\bigskip


\section{INTRODUCTION}
\label{intro}

Experimental studies of transport coefficients in systems of
strongly interacting atoms adsorbed on a crystalline surface show a
very rich and complicated behavior, especially as functions of the
atomic concentration. The variation
of the diffusion coefficient versus coverage is particularly
important for adsorbed layers where the concentration
may be varied in wide limits from zero (diffusion of isolated
adsorbed atoms) to very high values (for example, in some adsystems
the interatomic distance in a monolayer of adatoms is lower than
that in the corresponding massive crystal)~\cite{NV}. The
theoretical study of mass and charge transport in such systems is a
very difficult problem; however it was studied
for various kinds of interactions by Gomer et al~\cite{GOMER}
using Monte Carlo simulations. 
At high temperatures, transport coefficients
can be found with a perturbation technique starting from
the case of noninteracting atoms~\cite{diet}. At low temperatures,
the case of interacting atoms has been studied by a numerical
calculation of the transport properties of the standard
Frenkel-Kontorova (FK) model, which describes a chain of
harmonically interacting atoms subjected to a one-dimensional
sinusoidal external potential\cite{gillan2,gillan4}. Recently the
low-temperature behavior  of  a system of strongly interacting
atoms in a more general one-dimensional model has been
approximately treated  with a phenomenological approach which
introduces weakly interacting quasi-particles~\cite{BKlast}. This
method provides analytical estimates for the transport
coefficients, but it requires many approximations. In
particular the properties of the
quasi-particles involved in the theory are deduced from
results of the standard FK model which provides only a simplified
picture. Therefore it was necessary to check the validity of the
theoretical approach by numerical simulations of a model which is
sufficiently complicated to provide a reasonable description of a
real system. This is the aim of this paper which studies a
two-dimensional generalized FK model and also discusses some
experimental results in the same perspective.

The original FK model was introduced to analyze the dynamics
of dislocations in crystals~\cite{fk} by considering a
chain of interacting particles subjected to a
periodic substrate (on-site) potential.
It can describe, for example, a closely-packed row
of atoms in a crystal~\cite{pan},
a chain of atoms adsorbed
on stepped or  furrowed crystal surfaces~\cite{lnp},
a chain of ions in a ``channel''
of a  quasi-one-dimensional conductor~\cite{boyce},
hydrogen atoms in hydrogen-bonded systems~\cite{hb}, etc.
In all the cases mentioned above, the chain of interacting particles
is an intrinsic part of the whole physical system under consideration.
The role of the remainder of the system is played by an external
substrate potential and a thermal bath. Although it is still
oversimplified, the {\em generalized} FK model that we consider here
provides a rather complete description of a layer of atoms adsorbed
on a two-dimensional crystal surface. It includes
realistic (exponential)
interactions of particles instead of the harmonic springs of
the standard FK model, and the substrate
potential is three-dimensional. It is periodic in the two dimensions
parallel to the surface and has a Morse shape in the third direction,
orthogonal to the surface.

\bigskip
The transport properties of the system are described by
two coefficients, the mobility $B$ and the chemical
diffusivity $D_c$. The mobility defines the response of the
system to an infinitesimal d.c.\ force $F$,
\begin{equation}
J=\rho B F,
\label{B}
\end{equation}
where $J$ is the atomic flux caused by the force
and $\rho$ is the average atomic concentration.
On the other hand, the chemical diffusion coefficient $D_c$
connects the flux $J(x,t)$
in a nonequilibrium state to the gradient of the atomic
concentration when $\rho (x,t)$ slightly deviates from
its equilibrium value. According to Fick's law
\begin{equation}
\langle \langle  J(x,t) \rangle \rangle  \approx
- D_c \frac{\partial}{\partial x}
\langle \langle  \rho (x,t) \rangle \rangle ,
\label{Fick}
\end{equation}
where $\langle \langle \ldots \rangle \rangle $ stands
for the averaging over macroscopic distances $x \gg a_A$,
and $a_A$ is the average interatomic distance.
These two coefficients are coupled through the relation
\begin{equation}
D_c = {k_B T B\over \chi},
\label{DcB}
\end{equation}
where $k_B$ is Boltzmann's constant, $T$ the temperature and
$\chi$ the dimensionless susceptibility of the system.

The predictions of the phenomenological approach~\cite{BKlast} can
be summarized as follows. The mass transport is caused by kinks
which describe localized compressions or expansions of the chain and
therefore the mobility $B$ can be expected to be proportional to the
kink concentration. The kinks have two different origins,
``geometrical'' and thermal. We call geometrical kinks, the kinks
which result from the value of the coverage $\theta = N/M$, where
$N$ is the number of atoms and $M$ the number of wells of the
substrate potential on a given length. For $\theta = 1/q$, with
integer $q$ ($q = 1, 2, \ldots$), the system has a trivial ground
state with one atom at the bottom of the substrate wells every
$q^{\text{th}}$ well. When $\theta$ deviates slightly from such a
value, the difference is accommodated by the system by forming
localized discommensurations which are the geometrical kinks
(called also ``trivial kinks'' in the notation of
Ref.~\onlinecite{BKlast}). As the kink density increases when
$\theta$ deviates from $1/q$, the theory predicts that $B(\theta
)$  should exhibit local minima for any trivial ground state (GS)
of the system, such as
$\theta = 1$, $\theta =1/2$, $\theta =1/3$,  etc.  When $\theta =
p/q$ is a rational number with a larger numerator, such as $\theta
= 2/3$, the density of geometrical kinks becomes very large and one
could expect to get a high mobility $B$. The picture is however more
complicated because, due to their high density, the geometrical
kinks interact strongly and, when temperature is sufficiently low
with respect to their interaction energy, they tend to form a
regular lattice which is weakly pinned, giving a low mobility
for any rational $\theta$.  A slight deviation from
$\theta=p/q$ appears as discommensurations in the kink
lattice i.e.\ ``kinks in a kink lattice'', which are called
superkinks in Ref.~\onlinecite{BKlast}. These topological
excitations of the kink lattice contribute to mass transport
exactly as the trivial kink do, so that the mobility is expected to
exhibit local minima for
$\theta = p/q$ such as $\theta = 2/3$. In the limit
$T\rightarrow 0$ the function
$B(\theta )$ should therefore have minima at any rational $\theta $.
When temperature increases, the secondary minima disappear
because the kink lattice ``melts'' and moreover thermal
fluctuations create kink-antikink pairs which are thermally
activated. Consequently at high enough temperature the mobility is
expected to exhibit broad maxima between the primary minima at
$\theta = 1/q$. Such a behavior has been
observed in one dimensional models\cite{muna,gei,bishop79}.

The behavior of the diffusion coefficient $D_{c}$ is simpler than
the variation of $B(\theta)$ as predicted in Ref.~\onlinecite{bkz}.
According to Fick's law (\ref{Fick}), $D_c$ is the proportionality
coefficient between the (infinitesimal) gradient of the atomic
concentration and the flux of atoms caused by this gradient.
However, a gradient of atomic concentration automatically
produces a corresponding gradient of kink concentration. In the
standard FK model, where the elastic constant does not depend on
$\theta$ and where the parameters of kinks and antikinks are the
same,
$D_c(\theta )$ is the ratio of two quantities which vary similarly
so that it should be approximately constant and  coincide with the
kink (or antikink) diffusion coefficient. In the generalized FK
model the situation is different because the anharmonicity of the
interatomic interaction destroys the kink-antikink
symmetry~\cite{milmar}. The  effective interatomic
forces for a kink, which corresponds to a local contraction of the
chain, exceed those for an antikink which is associated to region of
a local extension. Thus, in comparison with an antikink, a kink is
characterized by a larger value of the rest energy and by lower
values of effective mass and activation energy for its
motion~\cite{bkz}. When the coverage passes through a commensurate
value $\theta_0$, the geometrical-kink density vanishes; for $\theta
< \theta_0$ the system has geometrical antikinks while for $\theta
> \theta_0$ the system has geometrical kinks.
 Therefore, when the coverage $\theta$ increases through a
commensurate value $\theta_0$, the activation energy for the chemical
diffusion should jump to a smaller value. Simultaneously
the value of $D_c$ should rise sharply when
the coverage $\theta $ exceeds the value $\theta_0$ that
characterizes a ``well-defined'' commensurate structure and one
could expect $D_c (\theta)$ to exhibit the shape of a Devil's
staircase. The abrupt
(jump-like) increase of $D_{c}(\theta )$ will only exist in the
$T\rightarrow 0$ limit and, for any $T\neq 0$, these jumps will
be smoothed owing to corrections from thermally excited kink-antikink
pairs.

\bigskip
In present paper we check these predictions by molecular dynamics
investigations of the low-temperature mobility and diffusivity
of a generalized FK model in  one and two dimensions.
In Sec.~\ref{model}, we
describe the model and define its parameters. Kink
parameters are calculated in Sec.~\ref{kinks}. Simulation
results for the mobility are presented in
Sec.~\ref{mobility}, and those for the chemical diffusivity
are described in Sec.~\ref{diffusion}. Sec.~\ref{Discussio}
discusses known experimental results in the framework of these
studies and Sec.~\ref{conclusion} concludes the paper.

\section{THE MODEL}
\label{model}

As for the standard FK model, we consider the dynamics of atoms
adsorbed on a periodic substrate. The displacement of
each atom is characterized by three variables:
$x$ and $y$
describe its motion parallel to the surface, while $z$ describes its
deviation orthogonal to the substrate. For the substrate potential,
we take the function
\begin{equation}
V_{sub}(x,y,z)=[V_{pr}(x;a_{sx},\varepsilon_{sx},s_{x})
              +V_{pr}(y;a_{sy},\varepsilon_{sy},s_{y})] e^{-\gamma'z}
              +V_z(z).
\label{sub}
\end{equation}
To model the substrate potential along the surface,
we use a deformable periodic potential which can be adjusted to
describe an actual crystal field\cite{peyrard-remoissenet},
\begin{equation}
V_{pr}(x;a_{sx},\varepsilon_{sx},s_x) = \frac{\varepsilon_{sx}}{2}
\frac {(1+s_x)^2[1-\cos (2\pi x/a_{sx})]}
{1+s_x^2-2s_x\cos (2\pi x/a_{sx})}.
\label{pr}
\end{equation}
Thus, $\varepsilon_{sx}$ corresponds to the activation
energy for diffusion of a single atom along the $x$
direction, $a_{sx}$ to the lattice constant and the
parameter $s_{x}$ ($|s_x|<1$) controls the shape of the
substrate potential. The frequency $\omega_x$ of a
single-atom vibration along the $x$ direction is connected
to the shape parameter $s_x$ by the relationship $\omega_x =
\omega_0 (1+s_x)/(1-s_x)$, where $\omega_0 \equiv
(\varepsilon_{sx}/2 m)^{1/2} (2 \pi/a_{sx})$ and $m$ is the
atomic mass. The potential
$V_{pr}(y;a_{sy},\varepsilon_{sy},s_{y})$ has the same form.

The potential perpendicular to the surface is modeled by the
Morse function
\begin{equation}
V_z(z) = \varepsilon_d \left( e^{-\gamma z}-1\right) ^2,
\label{Vz}
\end{equation}
which tends to the adsorption energy $\varepsilon_d$
when $z$ goes to infinity.
The anharmonicity parameter $\gamma$ is related to
the frequency $\omega_z$ of a single-atom vibration
in the normal direction by the relation
$\omega_z^2 = 2\gamma^2 \varepsilon_d /m$.

Finally, the exponential factor after the square brackets
of the right-hand side of Eq.~(\ref{sub}) takes into
account the decrease
of the influence of the surface corrugation
as the atoms move away from the surface,
so that $V_{sub}(x,y,z)\rightarrow \varepsilon_d$
when $z\rightarrow \infty$.

For the interaction between the atoms we take the
exponential repulsion
\begin{equation}
V_{int}(r) =  V_0 \exp (-\beta_0 r),
\label{int}
\end{equation}
where $V_0$ is the amplitude and
$\beta_0^{-1}$ determines the typical range of the interaction.
This potential is adapted to describe rather high coverages such
that the atoms interact through the repulsive branch of the
interatomic potential.
In numerical simulations, we can only include
the interaction of a given adatom with a finite number of neighbors.
Therefore, we use the standard approach of MD simulations and
introduce a cutoff distance
$r^{\ast}$.  We account only for the
interactions between the atoms separated by distances lower than
$r^{\ast}$ and
to reduce errors caused by this procedure,
the interaction potential (\ref{int}) is truncated as
\begin{equation}
\tilde{V}_{int}(r)=V_{int}(r) -
V_{int}(r^{\ast}) - V'_{int}(r^{\ast}) (r-r^{\ast}),
\label{intt}
\end{equation}
so that the interaction potential and force vanish
at the cutoff distance,
$\tilde{V}_{int}(r^{\ast})=\tilde{V}'_{int}(r^{\ast})=0$
(tilde will be omitted in what follows).
Besides, because we are using the repulsive interatomic interaction,
we have to fix the atomic concentration. It is imposed by
periodic boundary conditions in $x$ and $y$.
We place $N$
atoms in the fixed area $L_x\times L_y$,
$L_x=M_x a_{sx}$, $L_y=M_y a_{sy}$, so that
the dimensionless atomic concentration
is equal to $\theta=N/M$, where $M=M_x M_y$.

To model the energy exchange of the atoms with a thermal bath,
we use the Langevin equations for atomic coordinates $x_i$
\begin{equation}
m\ddot{x_i} + m \eta \dot{x_i} + \frac{d}{d x_i}
\left[
V_{sub}(x_i,y_i,z_i) + \sum_{j\neq i}
V_{int}(|\vec{r}_i-\vec{r}_j|)
\right]=F^{(x)} + \delta\!F_{i}^{(x)} (t),
\label{langevin}
\end{equation}
and similar equations for $y_i$ and $z_i$.
Here, $\eta$ corresponds to the rate of the energy exchange
with the substrate,
$\vec{F}=\{ F,0,0 \}$ to the dc driving force,
and $\delta\!F$ is a Gaussian random force
with correlation function
\begin{equation}
\langle \delta\!F^{(\alpha)}_i(t) \; \delta\!F^{(\beta)}_j (t') \rangle
= 2 \eta m k_B T
\delta_{\alpha \beta} \delta_{ij} \delta(t-t').
\label{random}
\end{equation}

We use a dimensional system of
units adapted to the scales of the problem. Distance is measured in
Angstr\"{o}ms, energy and temperature in eV. The
mass of an adatom is chosen as our mass unit ($m=1$). This imposes
a time scale. We measure time in units of the characteristic time
interval $t_0=2
\pi /\omega_x$. In the remainder of the paper, the measures of
other  dimensional physical quantities will be omitted,  but they
are all expressed in terms of the above-defined  units.

In order to be close to real physical systems, let us
take the adsystem {\bf K}-{\bf W}(112) as an example to
define the model parameters: the  {\bf W}(112) surface is
characterized by a strong anisotropy of the atomic
relief because it has close packed rows of substrate atoms
separated by furrows of atomic depth. Namely, in the
simulation, we put $a_{sx} = 2.74$~\AA\ and $a_{sy} =
4.47$~\AA\ which are respectively the distances between the
neighboring wells along and across the furrows on the {\bf
W}(112) surface, and  $\varepsilon_{sx} =
0.46$~eV and $\varepsilon_{sy} = 0.76$~eV for the
corresponding barriers (these values were taken
from Ref.~\onlinecite{parameter}). To model the shape of the
substrate potential, we have to define the parameters $s_x$ and
$s_y$. They can be estimated to be within the range
$[0.2,0.4]$~\cite{sxy}. For the sake of concreteness we took
$s_x=0.2$ and
$s_y=0.4$, which leads to the following frequencies of
adatom vibrations: $\omega_x=1.65$ and $\omega_y=2.02$. The
experimental value for the adsorption energy of {\bf K} on
{\bf W} is $\varepsilon_d =2.54$~eV~\cite{parameter}. For
the vibration frequency normal to the surface we took
$\omega_z=\frac{1}{2} (\omega_x + \omega_y) =1.84$,
which gives $\gamma =0.813$. For the interatomic
potential~(\ref{int}), we took the parameters $V_0=10$~eV
and $\beta_0=0.85$~\AA $^{-1}$. These choices give
reasonable values for adsystems~\cite{BM}: the
interaction energies between two adatoms occupying the
nearest wells along the furrow and across are equal to
$V_{int}(a_{sx})\approx 0.98$~eV and $V_{int}(a_{sy})\approx
0.22$~eV respectively. Finally, we have to define the rate
of energy exchange between the adatoms and substrate: we
took the typical value~\cite{fric} $\eta=0.1\; \omega_x=0.165$.
Note that although some of the  parameters are
chosen rather arbitrary, they are typical for metal atoms
adsorbed on metal substrates~\cite{zangwill}. However, as
the model is still oversimplified to
describe a real adsystem, we have to say that our choice of
parameters does not claim to provide a quantitative
interpretation of the  {\bf K}-{\bf W}(112) adsystem.
We do nevertheless believe on the qualitative description of the
effects under investigation and claim that typical adsystems on
anisotropic surfaces
(e.g. lithium and strontium on molybdenum~(112) surface, for
which experimental data on the detailed coverage dependencies
of diffusion characteristics are available~\cite{Vedula})
should exhibit a similar behavior.
Finally, for numerical solution of the Langevin equations~(\ref{langevin}),
 we use the standard fourth-order
Runge-Kutta method with the time step $\Delta t=t_0/20 =
0.19$, and the cutoff radius was taken as $r^{\ast}=2 a_{sy}
= 8.94$ \AA.

\section{KINKS}
\label{kinks}

As the kinks are the main objects of the phenomenological
approach~\cite{BKlast}, let us first calculate their
parameters. We recall that kinks can be defined for any
commensurate atomic structure $\theta_0=s/q$, where $s$ and
$q$ are relative prime integers; the kink (resp.
antikink) describes then the  minimally possible topologically
stable compression (resp. expansion) of the commensurate
structure. The kink is a quasiparticle, characterized by an
effective mass $m_k$, a rest energy $\varepsilon_k$, and the
Peierls-Nabarro (PN) amplitude $\varepsilon_{pn}$,
corresponding to the barrier for the kink translation along
the chain. These parameters are determined by the
dimensionless elastic constant $g_{eff}$ defined as
\begin{equation}
g_{eff}={a_{sx}^2\over 2 \pi^2 \varepsilon_{sx}}\
V''_{int} (a_{A}).
\label{geff}
\end{equation}

Analytically, the kink parameters may be found in the
low-coupling limit $g_{eff}\ll 1$ or, in the strong-coupling
(sine-Gordon) limit~\cite{BKlast} $g_{eff}\gg 1$, however,
usual real physical systems are characterized by the
elastic constant $g_{eff}\sim 1$, so that both
approximations are too crude to be applied to our case. For our
choice of model parameters, we have $g_{eff}\approx 0.6$ for
$\theta_0=1$. Therefore, we will calculate the kink parameters
numerically.

The numerical method was described in detail in previous papers
\cite{bkz,thierry}. Briefly, we have to choose first
an appropriate size of the
finite chain in order to insert a single kink into the
$\theta_0=s/q$ commensurate background structure;
the integers $N$ and $M$ must satisfy the
equation~\cite{bkz,classif}
\begin{equation}
q N = s M + \sigma\quad ,
\label{NM}
\end{equation}
where the kink topological charge $\sigma$ is equal to
$\sigma=+1$ for the kink and $\sigma=-1$ for the antikink.
In the simulation, we restrict ourselves to the concentration range
$[0.5,1]$ because for lower concentrations,
 the interatomic interaction is too weak and its effects would be
hardly observable, while at higher concentrations
the atoms begin to escape from the first adlayer
\cite{thierry,growth}.
As background structures, we chose the following coverages:
$\theta_0=1/2$, $\theta_0=3/5$, $\theta_0=2/3$,
$\theta_0=3/4$, $\theta_0=4/5$ and  $\theta_0=1$.
The corresponding values for the number of atoms $N_0$
and the number of minima of the substrate potential $M_0$
for the every $\theta_0$
are summarized in Table \ref{table1}.

We start with an appropriate initial configuration
and allow the atoms to reach the minimum-energy configuration
(see details in Refs.~\onlinecite{growth,zigzag}). This determines the
structure of a kink in its minimal energy state. Then, in order to
calculate the parameters that characterize the kink translation, we
choose a given atom in the kink region (see Ref.~\onlinecite{bkz}) and
move it to the right by small steps by imposing its $x$ coordinate
while all other degrees of freedom of the chain remain free
to adjust to every new position of the constrained atom.
This process allows us to find the saddle configuration and
therefore, the amplitude of the PN barrier $\varepsilon_{pn}$
as the difference of the saddle and GS energies.
Besides, the energy of creation of the kink-antikink pair
is determined as
$\varepsilon_{pair}=E\{\mbox{kink}[\theta_0]\}+
E\{\mbox{antikink}[\theta_0]\}-2 E\{\mbox{GS}[\theta_0]\}$,
where $E\{.\}$ is the energy of the corresponding configuration
(notice that it must satisfy the relation
$N\{\mbox{kink}[\theta_0]\}+N\{\mbox{antikink}[\theta_0]\}=
2 N\{\mbox{GS}[\theta_0]\}$ which is imposed by the total length
of the atomic chains along $x$). The kink parameters obtained by
this method are summarized in Table \ref{table1}, and the dependence
of the PN energy on the atomic concentration is presented in
Fig.~\ref{fig1}. Note that the function $\varepsilon_{pn} (\theta)$
has the shape of a devil's staircase~\cite{bkz}.

  From the kink parameters, the phenomenological
approach~\cite{BKlast} describes approximately
the low-temperature behavior of the system as follows.
For $\theta=1$ and $T \ll \varepsilon_{pair}\{\theta_0\}/k_B$
the concentration of thermally nucleated kink-antikink pairs
is equal to~\cite{seeger66,petuchov73,currie80,buttiker81,marchesoni91}
\begin{equation}
\langle \theta_{pair} \rangle \approx
C \exp (-\epsilon_{pair} /2 k_B T)\quad ,
\label{thetapair}
\end{equation}
where
$C \approx (2 \tilde{m}_k
\omega_x^2 a_{sx}^2 / \pi k_BT)^{1/2}$
and $\tilde{m}_k = (m_k m_{\bar{k}})^{1/2}$. For lower coverages
$\theta = \theta_0 = s/q$ Eq.~(\ref{thetapair}) should be properly
renormalized, which results in additional factor $1/q$ in its right-hand
side.

When the concentration $\theta$ slightly deviates from
the commensurate value $\theta_0$,
the thermal kinks are supplemented by the geometrical kinks
(if $\theta>\theta_0$) or antikinks (if $\theta<\theta_0$)
with a concentration $\theta_{geom}=q|\theta-\theta_0|$.
In the close vicinity of $\theta_0$, the total kink
concentration  can be found as
\begin{equation}
\langle \theta_{tot} \rangle \approx
\theta_{geom}+ 2 \langle \theta_{pair} \rangle.
\label{thetatot}
\end{equation}
The dimensionless susceptibility $\chi$
\begin{equation}
\chi =\langle \theta_{tot} \rangle /q^2 \theta
\label{chi}
\end{equation}
can then easily be obtained from Eqs.\
(\ref{thetapair}--\ref{thetatot}).

Let us now examine the phenomenology of consequently
melted kink superlattices\cite{BKlast}. In order to describe the
atomic mobility in terms of collective excitations, we must first
define the type of excitations that have to be considered. As an
example, let us select a concentration
$\theta$ in the neighboring of $\theta_0=2/3$.
For low $T$, $T<\varepsilon_{pair}\{2/3\}/2 k_B$,
we have to use the superkinks defined on the background
of the $\theta_0=2/3$ structure in the expressions
(\ref{thetapair}--\ref{chi}).
However, in the intermediate temperature range i.e.,
$\varepsilon_{pair}\{2/3\}/2 k_B < T <
\varepsilon_{pair}\{1/2\}/2 k_B$,
when the superkinks are destroyed by thermal fluctuations
while the trivial kinks (defined on the $\theta_0=1/2$ background)
are not yet destroyed, we have to substitute the parameters
of the trivial kinks in Eqs.~(\ref{thetapair}--\ref{thetatot}).
In particular, we should take $q=3$ for low $T$
but $q=2$ for intermediate temperatures.
In the latter case, however, the parameters of
trivial kinks may seriously differ from those calculated
for the ideal case, because the concentration of trivial kinks
at the $\theta=2/3$ coverage is very large and their interaction
is not negligible.

When the amplitude of the activation barrier for the kink motion
is known, the diffusion coefficient $D_k$
for a single-kink random walk
may be approximately calculated with the Kramers theory.
For $T<\varepsilon_{pn}/k_B$ this approach gives
\begin{equation}
D_k = D_{k0} \exp (- \varepsilon_{pn}/k_BT)\quad ,
\label{Dk}
\end{equation}
where
\begin{eqnarray}
D_{k0}
\approx \left\{ \begin{array}{lll} a^2 \omega_{pn}/2\pi, \;\;\,
& & {\rm if} \;\;\; \eta_{l} < \eta < \omega_{pn}, \\ a^2
\omega_{pn}^2 / 2 \pi \eta, \;\;\; & & {\rm if}
\;\;\; \eta > \omega_{pn}\quad .  \end{array} \right.
\end{eqnarray}
Here $\omega_{pn} \approx (\pi/a_{sx})(2 \varepsilon_{pn}/m)^{1/2}$,
$a=q\cdot a_{sx}$,
and $\eta_{l} = \omega_{pn}k_BT/2\pi \varepsilon_{pn}$.

If the interatomic interaction is strong enough,
the inequality $\varepsilon_{pn} < \varepsilon_{pair}$
may easily be fulfilled.
In this case, within the temperature interval
$\varepsilon_{pn} < k_BT < \varepsilon_{pair}$, the kink
diffusion coefficient is approximately equal to
(e.g. see Refs.~\onlinecite{remo,ber,march1})
\begin{equation}
D_k \approx \frac{k_BT}{m_k \eta} \left[ 1 -
\frac{1}{8} \left( \frac{\varepsilon_{pn}}{k_BT} \right)^2
\right]\quad .
\label{Dkk}
\end{equation}

Knowing the kink diffusion coefficient,
we can find the chemical diffusion coefficient
with the phenomenological approach~\cite{BKlast}
by the formula
\begin{equation}
D_c\approx {\langle \theta_k \rangle D_k +
\langle \theta_{\bar{k}} \rangle D_{\bar{k}}  \over
\langle \theta_{tot} \rangle}\quad  ,
\label{Dc}
\end{equation}
where for $\theta>\theta_0$ we should take
$\langle \theta_k \rangle = \theta_{geom} +
\langle \theta_{pair} \rangle $
and $\langle \theta_{\bar{k}} \rangle =\langle \theta_{pair} \rangle $,
while in the $\theta<\theta_0$ case, we have to substitute
$\langle \theta_k \rangle =\langle \theta_{pair} \rangle $
and $\langle \theta_{\bar{k}} \rangle =
\theta_{geom} +
\langle \theta_{pair} \rangle $.
Finally, the chain mobility may be found as $B=\chi D_c$.
These predictions should be now compared with the results of simulation
and it is the subject of the following section.

\section{MOBILITY}
\label{mobility}

To study the mobility, we use an algorithm where we look first for
the minimum-energy configuration of the system. Then, we increase
the temperature up to a given temperature $T$ by
small steps $\Delta T=T/50$ during the time $t_{therm}=100\; t_0 =
381$. At that point, we apply a small dc force $F=0.01$
which is gradually increased
 from $F=0$ to $F=0.01$ during the time $100\; t_0$,   and wait
during $t_{wait}=100\; t_0$ in order to allow the system to reach a
stationary state. Then, for the discrete times $t_n=n\; t_0$, we
measure the average velocity  $\langle v_{x}\rangle $ of the atoms
during $t_{run}=100\; t_0$, and finally, the procedure is repeated
$n_{ave}$ times ($n_{ave}=5$ in the simulation) with different
initializations of the random number generator in order to estimate
the error bars.

To demonstrate the effect of the transverse degrees of
freedom on the atomic mobility, we considered three different
cases of the generalized Frenkel-Kontorova model:

\begin{itemize}
\item a purely one-dimensional
atomic chain with atomic movement restricted to the
$x$ direction~(1D),

\item a quasi-one-dimensional atomic chain with two transverse
degrees of freedom $y$ and $z$ (we will call it the quasi-1D),

\item a true two-dimensional extension of the
FK model (2D).

\end{itemize}

Note that the interaction between the atoms  always has the
general form of Eq.~(\ref{int}) i.e.\ it has a  3D character
in the quasi-1D and 2D cases.

\bigskip
The quasi-1D case can be easily obtained from the general case by
choosing a period of one lattice constant in the $y$ direction. Namely, we put
$M_y=1$ so that all chains move in the same way and chose $N=5
N_0\{\theta \}$ and $M_x=5 M_0\{\theta \}$, where the integers
$N_0$ and
$M_0$ are taken from Table \ref{table1} for each coverage
$\theta$ in order to have 5 kinks or antikinks over the length
under investigation.

The results of the simulations are presented in Fig.~\ref{fig2}.
As expected from the phenomenological theory, at low temperatures
$B(\theta)$  does have local minima not only for the
trivial concentrations
$\theta=1/2$ and $\theta=1$, but also at the commensurate
concentrations $\theta=2/3$ and $\theta=3/4$. These two minima,
which involve a kink lattice,
disappear when the temperature is increased, while the minima for
the trivial  structures survive at any temperature. In the
simulation, minima do not appear for the other complicated GS
structures (e.g.
$\theta=3/5$  and $\theta=4/5$) because these higher
order structures correspond to too low ``melting'' temperatures of
the kink lattice.

In order to check completely the phenomenological
theory~\cite{BKlast}, it would be interesting
to see if these extra minima appear at very low
temperature, but the mobility is then too small to allow us to
obtain accurate results in a numerical simulation.
As the ``melting'' temperature is determined by the
magnitude of the effective elastic constant of the kink
lattice, one could attempt to increase the parameter
$V_0$ in Eq.~(\ref{int}).
But in that case, the repulsion between the atoms is too large
and they begin to escape from the minima of the substrate
potential in the direction orthogonal to the chain~\cite{zigzag}.
To prevent this escaping and allow the study of higher order
minima,  we artificially restricted the atomic
displacements to the $x$ dimension. This allowed us to take
$V_0=100$~eV.
This case corresponds to the so-called 1D case. The results are
shown on Fig.~\ref{fig3}.

\bigskip
As seen from Fig.~\ref{fig3}, in the 1D case the function
$B(\theta)$ has pronounced local minima at $\theta=2/3$ and
$\theta=3/4$ at much higher temperatures than in the quasi-1D case.
One can note also that the minimum at $\theta=3/4$ disappears when
the temperature is increased, while the minimum at $\theta=2/3$
survives in the whole range of investigated temperatures,
since it has a greater melting temperature.  However, the
minima at $\theta=3/5$ and $\theta=4/5$, where the kink
lattice has the period $5 a_{sx}$, are not found even for
this very strong interatomic repulsion. Note also that in
real physical systems, such as adsorbed layers, the
observation of local minima for these complicated structures
is unlikely due to existence of transverse degrees of
freedom.

In a one-dimensional model at high enough temperature
($T>\varepsilon_{pair}/k_B$), the mobility can be calculated
with  a perturbative approach starting from a system of
noninteracting atoms. The function $B(\theta)$ is given
by\cite{muna,gei,bishop79}
\begin{equation}
B \approx B_f \left\{
1 + \frac{1}{8} \left[ \frac
{(\varepsilon_{sx}/k_B T) \sinh (k_B T/\varepsilon_{sx} g_A)}
{\cosh(k_B T/\varepsilon_{sx} g_A) - \cos (2\pi a_A/ a_{sx})}
\right]^2 \right\}^{-1}\quad ,
\label{highT}
\end{equation}
where $B_f=1/m \eta$, $a_A=\theta a_{sx}$ is the average
interatomic distance and the elastic constant $g_A$ is
defined by  $g_A = a_{sx}^2 V_{int}^{\prime
\prime}(a_A)/2\pi^2  \varepsilon_{sx}$. Note that the
function (\ref{highT}) has local minima for trivial
configurations only, in agreement with the phenomenological
theory in the high temperature range. The chemical
diffusivity $D_c$ could be obtained replacing the  prefactor
$B_f$ in Eq.~(\ref{highT}) by
$a_{A}^2 V_{int}^{\prime \prime}(a_A)/m \eta$.
Figure~\ref{fig3} shows that Eq.~(\ref{highT})
describes the high-temperature simulation results
of the purely onedimensional FK model with a good accuracy (except
in the vicinity of the coverage $\theta=2/3$, where
$\varepsilon_{pair}/k_B > T$ even for the highest studied
temperature). For this model, the highest  possible
mobility $B_f =1/m \eta \approx 6.0 $, which corresponds to the
case of noninteracting free atoms, is reached in the  middle
of the interval $0.5< \theta <1.0$.

The high temperature mobility for the quasi-1D
chain with transverse degrees of freedom (curve (4) in
Fig.~\ref{fig2})  is approximately two times lower
than the values calculated with  Eq.~(\ref{highT}) for
corresponding parameters. Note that, for the chosen set of
parameters in a quasi-1D model, $T=0.05$~eV is the
highest temperature for which the determination of the
mobility {\em in the first monolayer of atoms} is possible. At
higher temperatures the atoms  start to escape from the first layer
to the second one, which may seriously distort the results (the
curve (4) in Fig.~\ref{fig2} is not plotted at $\theta> 0.8$ for
this reason). Even if one takes into account the fact that
the high temperature range required for the validity of
Eq.~(\ref{highT}) is not reached, the disagreement between
Eq.~(\ref{highT}) and the simulation results is large in
the quasi-1D case. This shows that the presence of
the transverse degrees of freedom has the same effect on the
mobility than an additional friction in the system. This can be
understood because some part of the work done by the external force
is used to excite the transverse degrees of freedom.

\bigskip
Finally, we also simulated the 2D Frenkel-Kontorova system
with $M_y=30$, $M_x=M_0\{\theta\}$ and $N=30 N_0\{\theta\}$.
The results, presented in Fig.~\ref{fig4},
show that there is no essential difference
between the $B(\theta)$ dependencies for the quasi-1D and 2D systems
except that 2D dependencies are systematically lower. The role of
the transverse degrees of freedom, already noticed for the
quasi-1D model, show up again here.  It is interesting to
notice that Fig.~\ref{fig4} shows for the 2D model at $T \le
0.01$~eV,
the  additional small minimum of $B(\theta)$ at $\theta=4/5$
predicted by the phenomenological theory, which reflects the
existence of the   kinks/antikinks on the background of this
coverage.

The plots of the mobility $B$ versus  inverse temperature shown in
Fig.~\ref{fig5} for the 2D model at  selected coverages  show that
the atomic mobility has an activated character in the investigated
range of temperatures and coverages.   The same qualitative behavior
was found for the 1D case. Using an Arrhenius form $B(T)=B_0 \exp
(-E_a/T)$, we can calculate the activation energy $E_a$ and the
prefactor $B_0$. Their dependence versus coverage
is shown in Fig.~\ref{fig6}.
The activation energy $E_a$ has a sharp maximum at the
coverage $\theta=2/3$ which corresponds to a well-defined
commensurate structure, while activation barriers on both
sides of $\theta=2/3$ are much lower due to the presence of
residual kinks/antikinks; the barrier for ``kink''
coverage $\theta=2/3 + \delta$ is lower than
that for ``antikink'' coverage $\theta=2/3 - \delta$. On
the other hand, the maxima of $E_a$ at the higher order
commensurabilities $\theta=3/4$ and
$\theta=4/5$ are much less pronounced. This is consistent with the
fact that these higher order commensurabilities hardly show up in
the mobility curves of Fig.~\ref{fig4}.

\bigskip
The main difference between the quasi-1D system and the true 2D
model is due to the interactions of kinks in the
nearest  neighboring channels.
For the repulsive interatomic
interaction studied in the present work, kinks in the nearest
neighboring channels repel each other at the $\theta=1$ coverage,
while for any $\theta <1$ they  attract each other and tend to
form domain walls.  With the short-range
(exponential) interaction studied in our simulations, two
kinks belonging to neighboring channels attract each other
with a potential $V_{kk}(x) \propto |x|$, contrary to the usual
law~\cite{lnp} $V_{kk}(x) \propto x^2$.  As a consequence,
the stiffness of the domain walls  vanishes and they can be
destroyed by thermal fluctuations or external forces for any $T\neq
0$ or $F\neq 0$. This is confirmed by the observation of snapshots
of the atomic configuration in the 2D model. During the time
evolution a domain wall of kinks is destroyed
as soon as the temperature is high enough to provide a
noticeable value of the mobility. For example  Fig.~\ref{fig7}
demonstrates the evolution of such a domain wall defined on the
background of the
$\theta =1/2$  commensurate structure. This case was chosen
because its kinks have the simplest structure and  are more visible
than kinks defined on the background of any other more complex
commensurate structure. The initially  well
defined kink wall (relaxed configuration for $T=0$) is
smeared out at $T=0.02$~eV and $F=0.01$,
although these values of temperature and external force
provide a very low value of the atomic  mobility ($B \approx
0.03$) at the chosen coverage. This instability of the kink domain
walls explains why the true 2D model give results which are not
fundamentally different from the results of the quasi-1D case.
Nevertheless, the interaction between atoms and kinks in the
nearest neighbor channels  does contribute to the dynamics of the
system. It results  in particular in the lowered values of
mobility for 2D case in  comparison with those for a 1D system.
However, in more realistic 2D models with long-range
interatomic forces such as elastic or dipole-dipole forces
due to the substrate, the role of the domain walls
might be more essential.

\section{DIFFUSION}
\label{diffusion}

The chemical diffusion coefficient is more difficult to calculate
by MD simulations than the mobility. It can be determined in
two ways. First, the susceptibility $\chi$ can be calculated with
one of the methods described in Ref.~\onlinecite{gillan4}
and then $D_c$ could be derived from the relation $D_c=B/\chi$.
However this approach relies on the accuracy of the two factors.
In the present work, we use a direct approach based on
the Fick law~(\ref{Fick}). We start from an nonuniform
initial concentration profile $\theta(x)$ and observe its
evolution  with time at a given temperature $T$   (we will
now use the notation $\theta$ instead of  $\langle \langle
\rho \rangle \rangle$ in the diffusion laws, assuming the
existence of local equilibrium). The variations
of the chemical diffusion coefficient $D_c$ with concentration
determine the diffusion profile~\cite{NV}. For instance, for an
approximately constant flux $J=-D_c \partial \theta/\partial
x$, flat sections of the observed concentration profile  (low
$\partial \theta / \partial x$) correspond to enhanced
diffusivity $D_c$, while sharp changes of concentration
within some concentration interval  (high $\partial \theta /
\partial x$)  indicate a lower diffusion coefficient $D_c$.

Quantitative data on the variation versus $\theta$ of the
diffusion coefficient $D_c$ can be obtained by studying
the concentration profiles given by the one-dimensional
diffusion equation
\begin{equation}
\frac{\partial }{\partial t} \theta(x,t) =
\frac{\partial}{\partial x}\left( D_c(\theta)
\frac{\partial \theta(x,t)}{\partial x}\right)\quad .
\label{difeq}
\end{equation}
The simplest case is the diffusion of an initial step-wise
profile in a spatially infinite system
which gives an explicit expression for the $D_c(\theta)$
function by the Boltzmann-Matano formula  (see
e.g. Ref.~\onlinecite{NV}). However, with periodic boundary
conditions, computational limitations do not allow us to
chose a period large enough to observe such a profile. Therefore, we
first derive an approximate expression of $D_c(\theta)$ using
the phenomenological equations
(\ref{thetapair}--\ref{Dc}), and then solve Eq.~(\ref{difeq})
with this $D_c(\theta)$ and periodic boundary conditions.
Finally, we compare the calculated profiles with those obtained from MD
simulation for the same initial distribution.

\bigskip
Let us first apply the kink-gas phenomenology to the
determination of $D_c$. We have
chosen the room temperature $T=0.025$~eV  ($290$~K)
since it divides the whole investigated
coverage interval $[0.5,1.0]$ into two parts which
differ by the mechanism of kink diffusion. For
the coverage range $0.5 < \theta < 0.66$ where the
condition $T<\varepsilon_{pn}/k_B$ is satisfied
(see Fig.~\ref{fig1}), the diffusion of kinks has an {\em
activated} character and, in terms of the Arrhenius
representation of the chemical diffusion coefficient $D_c=D_0
\exp (-E_{ac}/T)$, this means that the $D_c (\theta)$ dependence
is  determined mainly by the variations of the activation energy
(note that $E_{ac} \approx \varepsilon_{pn}$ according to
the Eq.~(\ref{Dk})). On the other hand, for  $0.66 <
\theta < 1.0$,  where $ T > \varepsilon_{pn}/k_B$ and where the
Peierls-Nabarro energy  $\varepsilon_{pn}$ shows only minor
changes with coverage, the mass
transport is carried  out by {\em free} diffusion of
kinks and the main  variations of chemical diffusion
coefficient will arise from the prefactor $D_0$.

One should keep in mind that the phenomenological equations
(\ref{thetapair}--\ref{Dc}) can be used only for those kinks which
are well defined as quasiparticles for a given temperature and a
given coverage interval, i.e.\ the condition  $T \ll
\varepsilon_{pair}/k_B$ must be satisfied. In other words,  the
concentration of thermally excited kink/antikink pairs
(\ref{thetapair}) for a given structure $\theta_0=s/q$  cannot
exceed the maximal possible value  $1/q$. For the present study, it
means that the quasiparticles which should be taken into account at
$T=0.025$~eV are the trivial kinks/antikinks of the trivial GS
$\theta_0=1/2$ and $\theta_0=1$ and superkinks defined on the
background of
$\theta_0=2/3$ structure.  We pointed out in Sec.~\ref{kinks} that,
for our model parameters, accurate kink parameters can only be
determined numerically. But, in order to solve Eq.~(\ref{difeq})
we need some expression for the kink masses.  Since
the maximum value of the dimensionless elastic constant
$g_{eff}$ defined by Eq.~(\ref{geff}) approximately equals
$0.6$ for the chosen set of model parameters, the best estimate
is given by the low-coupling limit expression~\cite{BKlast}  $m_k
\approx m_{\bar{k}} \approx m/q^2$.

Theoretically, the application of the Eqs.
(\ref{thetapair}--\ref{Dc}) for the determination of $D_c$ is only
strictly valid in the close vicinity of the
commensurate structures, where the concentration of residual kinks
is low   (in our case, near the coverages $\theta_0=1/2$,
$\theta_0=2/3$ and  $\theta_0=1$). Since we need $D_c(\theta)$ for
all intermediate $\theta$ values, we have to interpolate between
these specific $\theta$ values. We calculated the values of
$D_c$ (indicated by the plus signs in Fig.~\ref{fig8}) up to the
middle points between these specific coverages  and used a weighting
coefficient, plotted in inset in Fig.~\ref{fig8}, to mark the
significance of each point in the  subsequent interpolation
procedure. The final form of  $D_c(\theta)$ is taken as a
superposition of
$\tanh(\theta)$ functions chosen to provide a good fit of the value
deduced from the phenomenological theory around the coverages
$\theta_0=1/2$, $\theta_0=2/3$ and  $\theta_0=1$ where it is
accurate. Although this procedure cannot avoid some arbitrariness,
we keep it to a minimum by putting the weighting factor to zero
wherever the theoretical formula for $D_c(\theta)$ is not valid.
The general shape of the  interpolated  $D_c(\theta)$
reflect the expected general variations of the chemical
diffusivity versus coverage. We see that this function
monotonically increases in the  region $1/2 < \theta < 2/3$
(corresponding to the decrease of the activation energy
$E_{ac}$ which can be deduced from Fig.~\ref{fig1}), while at higher
coverages
$D_c$ start to decrease. The {\em high-temperature} behavior of
$D_c$ at high coverages is given in the kink-gas approach by the
variation of kink mass. It can be also interpreted in terms of
the Arrhenius formula  $D_c=D_0 \exp (-E_{ac}/T)$. Generally,
 $D_0$ and $E_{ac}$  change in a similar manner
(so-called {\em compensation effect}~\cite{NV}). At high
enough temperatures, the slow decrease of the $E_{ac}$  at
$\theta > 2/3$ (see  Fig.~\ref{fig1}) leads only to a slow change
of the exponential term of the Arrhenius formula.
The fast drop of $D_c$ must thus be attributed to
the prefactor $D_0$.

\bigskip
Once $D_c$ is known, the second step is to solve
Eq.~(\ref{difeq}) with this
$D_c$ dependence and compare with the MD simulations.
To deduce a local coverage from the MD atomic configuration at a
given time
$t$, we calculate the occupation numbers $n(i_x,i_y;t)$
(where $i_x=1,\ldots,M_x$ and $i_y=1,\ldots,M_y$) defined as
the number of atoms in the given elementary cell
$(i_x,i_y)$. The results of the simulations for the 2D model
($M_x=84$, $M_y=60$ and $N=3780$) are presented in
Fig.~\ref{fig9}. We started with an artificially prepared
step-wise initial configuration: concentration $\theta=1$
in the central region of the lattice (for
$i_x= M_x/4+1,\ldots,3 M_x/4$) and $\theta=0.5$ outside of this
region. The system is then
allowed to evolve according to the Langevin
equations~(\ref{langevin}).  The concentration
profile $\theta(i_x,t_n)=\sum_{i_y=1}^{M_y}
n(i_x,i_y;t_n)/M_y$ is recorded at dates $t_n=n t_0$. The
simulation was repeated five times in order to average the profiles
and decrease statistical fluctuations. The simulation profiles for
different times are shown in Fig.~\ref{fig9} with symbols. They
have a flatter section in the middle of the studied
coverage interval, corresponding to the coverage region with
enhanced diffusivity. Moreover, in the same figure, we plot
with solid lines the theoretical profiles obtained from the
numerical solution of the diffusion equation (\ref{difeq})
with $D_c(\theta)$ plotted in Fig.~\ref{fig8}.
The data presented in Fig.~\ref{fig9} show that
the theoretical and simulation  results are in a very good
agreement which validates the phenomenological approach
used for determining the diffusion coefficient.


\section{DISCUSSION IN RELATION WITH EXPERIMENTS}
\label{Discussio}

It is important to examine the applicability
of the theoretical results to real physical systems such as
atomic layers adsorbed on crystal surfaces. Although experiments
cannot provide results as detailed as the numerical simulations, a
comparison is possible. Our model is oversimplified to describe
quantitatively a real adsystem  --- although we chose some model
parameters  close to those available for the {\bf K}-{\bf W}(112)
adsystem ---, mainly because the interatomic interaction in real
adsystems is much more complicated than the exponential interaction
used in the present work\cite{BM}. In the case of
adsorption on isotropic surfaces, diffusion is affected by the
formation of domain walls, especially when the interatomic
interaction is long-range. Our results are more suitable to
describe highly anisotropic surfaces for which the interaction
between neighboring channels is sufficiently weak to reduce the
role of two-dimensional domain walls. We obtain a
qualitative agreement with experiments on diffusion of atoms
adsorbed on highly-anisotropic furrowed surfaces.

There are very few experimental data on the
variation versus coverage of the diffusion coefficient
for atoms adsorbed on highly anisotropic (furrowed)
surfaces.  Some data have however been obtained using the
field emission fluctuations method \cite{KW112} for the {\bf
K}-{\bf W}(112) and
the  diffusivity was found to increase strongly in the
region of the commensurate-incommensurate transition: this was
interpreted in terms of fast diffusion of {\em solitons}
\cite{KW112}.

Detailed
dependencies $D_c(\theta)$ and $E_{ac}(\theta)$ in the wide
coverage interval $[0.05,1.5]$ are available~\cite{Vedula}
for the {\bf Li}-{\bf Mo}(112), where the
interaction between {\bf Li} adatoms on {\bf Mo}(112) is
long-range and anisotropic. Besides the short range forces, the
interaction between the adatoms includes also a dipole-dipole
repulsion and an oscillating part due to substrate-mediated electron
exchange~\cite{BM}. This is responsible for the existence of peculiar
chain-like structures p(1$\times$4) and p(1$\times$2), formed  by
first-order transitions, for coverages $\theta < 0.5$
\cite{GM}. In this range of $\theta$, the diffusivity $D_c$ was
found to depend only weakly on coverage. At higher
coverages $\theta > 0.5$,  the repulsion  between  {\bf Li}
adatoms starts to play a larger role. This
results first in the formation of one-dimensional incoherent
structures at  $\theta \approx 2/3$, and then the adlayer
exhibits a one-dimensional compression along the direction of
furrows \cite{GM}. In this coverage range,
$D_c(\theta)$ was found to increase strongly and monotonically
with coverage at low  temperatures.
The sharpest increases of diffusivity at low
temperatures ($T<250K$) appear for the commensurate
coverages  $\theta =2/3$ and $\theta =1$. This behavior of $D_c$
coincides qualitatively with the predictions of the kink-gas
approach \cite{BKlast} and our numerical simulations.
Moreover the activation
energy $E_{ac}$ for chemical diffusion, obtained in
Ref.~\onlinecite{Vedula} from the slopes of Arrhenius plots of
$D_c$, exhibits a monotonical decrease as coverage increases
(except for small maxima at coverages slightly above the
commensurate values $\theta =2/3$ and $\theta =1$), which  is close
to the behavior of
 $\varepsilon_{pn}$ in Fig.~\ref{fig1}. It is also interesting
to notice that if $D_c(\theta)$ is plotted at temperatures higher
than 300K from the  values of prefactor the $D_0$ and
activation energy $E_{ac}$ measured in Ref.~\onlinecite{Vedula},
it shows a nonmonotonic behavior with a minimum around
$\theta=1$, which is very similar to the behavior of $D_c$
at $T=300K$ in the two-dimensional FK
model considered here (see section~\ref{diffusion}).

Finally, preliminary results of the diffusion study in
{\bf Sr}-{\bf Mo}(112)  system \cite{Vedula} have demonstrated, that
diffusivity increases sharply at coverage $\theta \approx
0.5$ which correspond to the commensurate  (4$\times$2)
structure of strontium atoms (while at higher coverages {\bf
Sr} atoms form incommensurate structures) . One may
speculate that this enhanced diffusivity is
provided by the fast kink diffusion.

\section{CONCLUSION}
\label{conclusion}

The aim of this paper was to check the predictions of the
kink-gas approach with the help of Molecular Dynamics simulations.
The validity of the kink gas approach ought to be questioned
because it is based on one-dimensional models while real adsystems
are 2D (or even 3D taking into account the possible motion of
adatoms orthogonal to the surface).

First, we compared the kink-gas
approach [Eqs.~(\ref{thetapair}--\ref{Dc})] and high-temperature
formula (\ref{highT}) for  mobility
$B$ and chemical diffusivity $D_c=k_B T B/\chi$ with the
results of our simulation of FK models with transverse
degrees of freedom. We have
found only a {\em qualitative} agreement between the  $B(\theta)$
dependencies obtained in MD simulation of 2D and 1D model
with transverse degrees of freedom and those predicted  by
kink-gas approach.
In section~\ref{mobility}, we showed that  the mobility
$B$ is strongly reduced when additional transverse
dimensions are involved in the system. A {\em
quantitative}  estimation of $B$  using
Eqs.~(\ref{thetapair}--\ref{Dc}) shows that the kink-gas approach
overestimates significantly the mobility
unless we artificially introduce a higher effective friction
$\eta_{eff}$. This can be understood because some part of
energy brought into the system by the external
force is absorbed by extra degrees of freedom.  For example,
in the simplest case of harmonically interacting atoms at high
temperatures, the  mobility has to be renormalized by a factor
$1/3$ (i.e. $\eta_{eff}=3\eta$) due to the presence of  2
additional degrees of freedom, since the  energy is
redistributed uniformly ($\sim k_B T/2$) between all three
degrees of freedom. But in our case, where interactions
between atoms is anharmonic and mobility is investigated
at low  temperatures, a reliable determination of
effective friction $\eta_{eff}$ is not possible.

By contrast, the study of chemical diffusion
(Sec.~\ref{diffusion}) demonstrates  {\em qualitative} and
{\em quantitative} agreement between MD simulation data and
the kink-gas approach. The reason is that in the case of
thermal diffusion, the chemical diffusion coefficient
$D_c=k_B T B/\chi$ does not change with additional degrees
of freedom since the ``external force'' is provided by the
thermal energy of the system and is proportional to the
gradient of chemical potential~$\mu$. As the Fick law
(\ref{Fick}) may be rewritten  as  $J=\rho B \nabla \mu =
\rho B (\partial \mu / \partial \rho)  \nabla  \rho = k_B T
B/ \chi \nabla \rho$,  where $\chi \equiv (\partial \ln
\rho)/(\partial \mu /k_B T)$, it is clear that  an
additional transverse dimension to the system leads simultaneously
to an increase of free energy and chemical potential $\mu$. In
other words, the decrease of system's mobility  due to
an extra transverse dimension of the system is
compensated by a corresponding decrease of susceptibility
$\chi$ of the system, so that $D_c$ remains approximately
the same.

Our results
allow us to conclude that the phenomenological kink-gas approach
provides a good qualitative explanation not only for Molecular
Dynamics simulation data of the mobility and diffusivity versus
atomic coverage in the  generalized 2D Frenkel Kontorova model, but
also for some experimental results on the coverage
dependence of surface diffusion. Obviously, for a better
description of the real adsorption systems, the Frenkel
Kontorova model should take into account long-range,
anisotropic, realistic interatomic interaction and the
presence of surface defects.

\acknowledgments
O.~M.~B. and M.~V.~P. were partially supported by Grant
K5T100 from the Joint Fund of the Government of Ukraine and
International Science Foundation. The work of M.~V.~P. was
also supported in part by NATO Linkage Grant LG 930236.
M.~V.~P. gratefully acknowledges the hospitality  of the
Ecole Normale Sup\'{e}rieure de Lyon,  where an essential  part
of this work has been done.


\newpage
\begin{table}
\caption{Parameters of kinks: $N_0$ number of atoms, $M_0$ number
of minima of the substrate potential for one period of the system
along $x$; $\varepsilon_{pair}$ creation energy of a kink-antikink
pair and $\varepsilon_{pn}$ amplitude of the Peierls-Nabarro
potential.}
\bigskip

\begin{tabular}{ccccccccc}
structure & $N_0$ & $M_0$ & $\varepsilon_{pair}$ (eV)
                             & $\varepsilon_{pn}$ (eV) \\
\tableline
antikink[1/2]     & 21 & 43 & --- & 0.378  \\
$\theta_0=1/2$    & 21 & 42 & 0.759  & --- \\
kink[1/2]         & 21 & 41 & --- & 0.0849 \\
\tableline
antikink[3/5]     & 22 & 37 & --- & 0.0848 \\
$\theta_0=3/5$    & 21 & 35 & 0.007  & --- \\
kink[3/5]         & 20 & 33 & --- & 0.0813 \\
\tableline
antikink[2/3]     & 21 & 32 & --- & 0.0812 \\
$\theta_0=2/3$    & 20 & 30 & 0.170  & --- \\
kink[2/3]         & 21 & 31 & --- & 0.0192 \\
\tableline
antikink[3/4]     & 20 & 27 & --- & 0.0184 \\
$\theta_0=3/4$    & 21 & 28 & 0.055  & --- \\
kink[3/4]         & 22 & 29 & --- & 0.0087 \\
\tableline
antikink[4/5]     & 19 & 24 & --- & 0.0086 \\
$\theta_0=4/5$    & 20 & 25 & 0.018  & --- \\
kink[4/5]         & 21 & 26 & --- & 0.0071 \\
\tableline
antikink[1]       & 21 & 22 & --- & 0.0071 \\
\end{tabular}
\label{table1}
\end{table}

\begin{figure}
\caption{Peierls-Nabarro energy
versus coverage. Coverages corresponding to simplest commensurate
structures are shown with dashed lines.}
\label{fig1}
\end{figure}

\begin{figure}
\caption{The mobility $B$ of the quasi-1D FK model
with transverse degrees of freedom
as a function of the coverage $\theta$
at selected temperatures
$T=0.0025$~eV [curve (1)],
$T=0.005$~eV (2),
$T=0.020$~eV (3),
$T=0.050$~eV (4).}
\label{fig2}
\end{figure}

\begin{figure}
\caption{The mobility $B$  versus coverage $\theta$
for purely 1D model with $V_0=100$~eV at different
temperatures $T=0.005$~eV (diamonds and dotted line),
$T=0.05$~eV (asterisks  and dashed line), $T=0.10$~eV
(triangles and solid line). For a better presentation, the data for
the two lower temperatures are plotted only within $0.72 < \theta < 0.8$,
because at other coverages they are the same as for $T=0.10$~eV.
The dash-triple-dotted line is the Eq.~(\protect\ref{highT})
for $T=0.10$~eV.}
\label{fig3}
\end{figure}

\begin{figure}
\caption{The mobility $B$ versus coverage $\theta$ for
2D model at selected temperatures
$T=0.005$~eV [curve (1)],
$T=0.010$~eV (2),
$T=0.020$~eV (3),
$T=0.030$~eV (4),
$T=0.050$~eV (5).}
\label{fig4}
\end{figure}

\begin{figure}
\caption{The mobility $B$ of the two-dimensional model
versus the inverse temperature:
(a) for coverages $\theta = 0.51$ (diamonds) and $\theta = 0.60$ (triangles),
(b) kinks (squares), antikinks (diamonds) and
the background commensurate structure (triangles)
for $\theta_0=2/3$.
(c) for coverages $\theta = 0.72$ (diamonds), $\theta = 0.80$
(triangles).}
\label{fig5}
\end{figure}

\begin{figure}
\caption{The activation energy $E_a$ and prefactor $B_0$
for the mobility versus coverage $\theta$
in the case of the two-dimensional FK model.}
\label{fig6}
\end{figure}

\begin{figure}
\caption{Snapshot pictures for two-dimensional model at $\theta = 0.51$.
(a) The initial relaxed  configuration ($T=0$),
(b) The atomic configuration
after the evolution time $t \approx 1600$ at $T=0.02$~eV}
\label{fig7}
\end{figure}

\begin{figure}
\caption{Chemical diffusion coefficient $D_c$ versus
coverage for $T=0.025$~eV. Crosses correspond to the $D_c$
values calculated with  phenomenological equations
(\protect\ref{thetapair}--\protect\ref{Dc}); the full curve
corresponds to the $D_c(\theta )$ dependence, interpolated
with the help of weighting coefficient presented in the
inset.}
\label{fig8}
\end{figure}

\begin{figure}
\caption{Evolution of the coverage profile versus time:
$t=0$~(a), $t=190$~(b), $t=763$~(c) and $t=1715$~(d).
The triangles correspond to the simulation of
the 2D model with $M_x=84$, $M_y=60$ and $N=3780$
at room temperature $T=0.025$~eV,
and the full curves are the solution of the
diffusion equation (\protect\ref{difeq}). Coordinate $x$ is indicated in
lattice units $a_{sx}$}
\label{fig9}
\end{figure}
\end{document}